\documentclass[preprint,review,12pt]{elsarticle}

\usepackage{amssymb}

\usepackage[nodots,nocompress]{numcompress}

\biboptions{super}

\journal{Nuclear Physics B}

\begin{document}

\begin{frontmatter}



\title{Rashba coupling induced spin susceptibility and magnetic phase transition of conduction electrons in monolayer graphene }


\author{Fatemeh Delkhosh
}
\author{A. Phirouznia
}
\address{Department of Physics, Azarbayjan Shahid Madani University
, 53714-161, Tabriz, Iran}
\cortext[cor1]{Corresponding author.} \ead{Phirouznia@azaruniv.ac.ir}
\begin{abstract}
Using the Kubo formalism, the magnetic properties of the system in the linear
regime have been investigated. Mainly the effect of non-magnetic substrate on the spin susceptibility is calculated.
Results show that the Rashba coupling interaction
has a central role in the magnetic response function of the system and it is really remarkable since this type of spin orbit coupling can be effectively controlled by an external gate voltage. Most importantly it was shown that, in the presence of the Rashba interaction a magnetic phase transition could be observed. This magnetic phase corresponds to a magnetic order of conduction electrons that takes place at some especial frequency of external magnetic field.
\end{abstract}


\end{frontmatter}


\section{Introduction}
\label{intro}
One of the most fascinating discoveries in condensed matter physics is graphene. The nature of the unusual  quantum Hall effect  and   the quasi relativistic energy spectrum of Dirac fermions  in graphene   has attracted more and more attention.
\\
Graphene is a single layer of a hexagonal lattice of carbon atoms. Crystal structure of the honeycomb lattice of graphene contains two different sublattices \cite{Geim,Kim,Wallace}. Graphene can  be  accumulated in the 0D fullerene, the rolling 1D nanotubes, put together, it is converted to graphite\cite{Geim}. Graphene, due to its structure has many unique features. It has been proved that graphene is the strongest material ever measured. Hexagonal unit cell of graphene has two carbon atoms and it is a cross-sectional area equal to 0.052 square nanometers. Graphene is almost fully transparent only 3.2 percent of the light is absorbed by the optical domain. This number indicates that suspended graphene does not have any color.
\\
Among the interesting physical properties of the mono-layer graphene, spin-dependent characteristics of the graphene has attracted tremendous attention due to its peculiar physical properties and huge potential in spin-dependent device applications.
Spin is the unique inner degree of freedom of the electron. Employing the spin of electron in electronic and optoelectronic devices could propose new features. This could be considered as the ambitious goal of spintronics. It is included concepts exciting and challenging, but also is sometimes questionable. Non-equilibrium spin-injection and control skills of polarization in a given system, have a fundamental role in the modern trend of semiconductor spintronics. spin-orbit coupling based on both objective is to achieve. Electrical spin injection and manipulating spin polarization could be achieved by Rashba coupling based technologies.
\\
This kind of spin-orbit interaction arises due to the lack of structural inversion symmetry \cite{Kane}. The importance of this mechanism lies in the fact that the asymmetry of the confining potential can be changed by electrostatic methods \cite{Yu}. The Rashba spin-orbit interaction can be changed by an external gate voltage. The Rashba interaction removes the spin degeneracy of conduction band even in the absence of an external magnetic field.
\\
Graphene could display a prominent role in the filed of spintronics. This is due to the fact that the Rashba coupling strength could be relatively high in graphene. The Rashba interaction strength can be increased up to 200 $meV$. This can be realized by direct growth of graphene on a ferromagnetic substrate using the catalyzed methods \cite{M,Emmanuel}. It should be noted that isolated graphene cannot be obtained by the known synthesis processes and it should be placed on a substrate. Therefore the Rashba coupling in graphene could also be induced by this substrate. The Rashba spin-orbit interaction can be adjusted by the gate voltage and the type of substrate. The effect of the displacement and dispersion caused by the Rashba coupling is expected. In particular, the Rashba spin-orbit interaction, results in a energy gap at Dirac points and gives a limited mass to the carriers in the graphene sheet \cite{M}. As long as the Rashba interaction is stronger than the intrinsic spin-orbit interaction, energy bands near the Dirac points of the Brillouin zone will show trigonal warping \cite{P}.
\\
Susceptibility using the free electron model to the electric field induced by the spin density is described \cite{Sigurdur,Catalina}. Meanwhile the Rashba coupling plays a major role in spin susceptibility of the free electron system. Then the spin current of the system could be obtained by using the continuity equation for spin density.  In the current work spin susceptibility of the mono-layer graphene has been obtained in the presence of the Rashba spin-orbit coupling. Regarding the high magnitude of the Rashba coupling strength in graphene and its linear dispersion relation. Spin susceptibility of the mono-layer graphene could be quite different from that of obtained for two-dimensional electron gas system.
\section{Model}
It should be noticed that the general form of the effective Hamiltonian of graphene at Dirac point approximation in the presence of Rashba coupling is such as:
\begin{equation}
\label{e2}
H=\gamma(\sigma_x \xi_z k_x +\sigma_y k_y)+\frac{1}{2}\lambda(\sigma_x s_y - \sigma_y\xi_z s_x)
\end{equation}
where $\gamma=\hbar v_f$ and $v_f\simeq 10^6 m/s$ Fermi velocity of graphene, $\sigma$ is the Pauli matrix of pseudospin and s is the  spin operator. In addition $\lambda$ is the Rashba coupling strength and $\xi_z=\pm 1$ defines the $K$ and $K^\prime$ Dirac points respectively.  It was also assumed that $\hbar/2$ is the unit of the spin.
\\
Hamiltonian of the mono-layer graphene in the presence of the Rashba spin-orbit coupling at the K point, can be obtained in the four-component spinor $\Psi$
basis, as follows:
\\
\begin{eqnarray}
\mathbf{H} = \left(
\begin{array}{cccc}
0 & 0 & \gamma k_- & 0 \\
0 & 0 & i\lambda & \gamma k_-  \\
\gamma k_+  & -i\lambda  & 0 & 0\\
0 &\gamma k_+  & 0 & 0
\end{array} \right)~~~~
\mathbf{\Psi} = \left(
\begin{array}{c}
a^\dag_{k\uparrow}  \\
a^\dag_{k\downarrow}\\
 b^\dag_{k\uparrow}\\
b^\dag_{k\downarrow}
\end{array} \right)
\end{eqnarray}\\
Note that $k_\pm=k_x \pm ik_y$, $a^\dag_{ks}$ and $b^\dag_{ks}$ are creation operators of an electron with given spin, $s$, in $A$ and $B$ sublattices respectively. In the other word the on-site spin-dependent Bloch functions ($|\psi_{ak}\rangle\otimes|s\rangle$ and $|\psi_{bk}\rangle\otimes|s\rangle$) have been chosen as a representation basis.
\\
The eigenvalues of K-point Hamiltonian can be obtained as:
\begin{eqnarray}
E_{\nu\mu}&=&\frac{\nu\mu}{2}(\sqrt{\lambda^2+4\gamma^2 k^2}-\mu\lambda)
\\
&=&\frac{\nu\mu}{2}(\Lambda_k-\mu\lambda)
\end{eqnarray}\\
In which in the above expression $ \nu,\mu =\pm 1$ and normalized eigenvectors are as follows:\\
\begin{equation}
\mathbf{|\nu\mu >} =B_k\left(
\begin{array}{c}
\mu \frac{ik_-}{k_+}\\
\\
-\frac{\nu\lambda -\nu\mu \Lambda_k}{2\gamma k_+}\\
\\
\frac{-i(\lambda - \Lambda_k)}{2\gamma k_+}\\
\\
1
\end{array} \right);
\end{equation}\\
for $ \nu=\pm,\mu =+ 1$ and
\begin{equation}
\mathbf{|\nu\mu >}=A_k\left(
\begin{array}{c}
\mu \frac{ik_-}{k_+}\\
\\
-\frac{\nu\lambda -\nu\mu \Lambda_k}{2\gamma k_+}\\
\\
\frac{-i(\lambda + \Lambda_k)}{2\gamma k_+}\\
\\
1
\end{array} \right)
\end{equation}
for $ \nu=\pm,\mu =- 1$,
where $A_k =\frac{\gamma k}{\sqrt{\Lambda_k ^2 +\lambda\Lambda_k}}~~ B_k=\frac{\gamma k}{\sqrt{\Lambda_k ^2 -\lambda \Lambda_k}}~~\Lambda_k =\sqrt{\lambda ^2+4\gamma ^2 k
}$.
\\
The spin susceptibility of the mon-olayer graphene via the linear response formalism kubo is calculated as follows\cite{Catalina}:\\
\begin{eqnarray}
\chi_{\mu\mu^\prime}(\omega)&=&\frac{i}{\hbar}\int dt e^{i(\omega + i\eta)}<[s_\mu(t) , s_{\mu^{\prime}}(0)]>~~
\mu, \mu^{\prime}=x,y\\
  <[s_\mu(t) , s_{\mu^{\prime}}(0)]>&=&\int \sum_{\lambda} \frac{k dk d\theta}{e^\frac{\varepsilon_{\lambda}(k)-\mu}{k_BT}+1}<\lambda|[s_\mu(t) , s_{\mu^{\prime}}(0)]|\lambda>
\end{eqnarray}
In the above equation$<...>=\sum_{\lambda} \int d^2k f(\varepsilon_{\lambda}(k)) {<...>}$
represents the thermal average. where k Fermi wave vector, $f(\varepsilon_{\lambda}(k))$ Fermi distribution function,  $\omega$ Frequency, $k_B$ Boltzmann constant,T Temperature and $\eta\rightarrow 0^{+}$. According to the Heisenberg picture $s_i(t)= e^{\frac{iHt}{\hbar}}s(0)_{i}e^{\frac{-iHt}{\hbar}}$, $ <\nu\mu |[s_i(t) ,~ s_j(0)]|\nu\mu >$ is obtained:
\begin{eqnarray}
<\nu\mu |[s_i(t) , s_j(0)]|\nu\mu >&=&\sum_{\nu^\prime\mu^\prime}2i \times Im[ e^{\frac{ i(E_{\nu\mu}-E_{\nu^\prime\mu^\prime})t}{\hbar}}\nonumber
\\
&&\times <\nu\mu | s_i(0)|\nu^\prime \mu^\prime ><\nu^\prime \mu^\prime |s_j(0)|\nu\mu >].
\end{eqnarray}
\\
Where the matrix elements of $s_x$ and $s_y$ on the basis of $|\nu\mu>$ is given as:
\begin{equation}
\textbf{s}_x = \left(
\begin{array}{cccc}
C_k(\lambda - \Lambda_k ) &D_k\lambda & 0 & -F_k\Lambda_k\\
\\
D_k\lambda & G_k(\lambda +\Lambda_k ) & F_k\Lambda_k & 0\\
\\
0 & -F_k\Lambda_k & C_k(\lambda - \Lambda_k ) & D_k\lambda \\
\\
F_k\Lambda_k & 0 & D_k\lambda & G_k(\lambda +\Lambda_k )
\end{array} \right),
\end{equation}
\\
In which $\; C_k=\frac{-B_k^2}{\gamma k^2}(2k_y)\; , D_k=\frac{-A_k B_k}{\gamma k^2}(2k_y),\; F_k= \frac{A_k B_k}{\gamma k^2}(2ik_x)$ ,\\
\\
$G_k=\frac{-A_k^2}{\gamma k^2}(2k_y)$
\\
\\
\begin{equation}
\textbf{s}_y = \left(
\begin{array}{cccc}
C_k^{'}(\lambda - \Lambda_k ) & D_k^{'}\lambda & 0 & -F_k^{'}\Lambda_k\\
\\
D_k^{'}\lambda & G_k^{'}(\lambda +\Lambda_k ) & F_k^{'}\lambda & 0\\
\\
0 & -F_k^{'}\Lambda_k & C_k^{'}(\lambda - \Lambda_k ) & D_k^{'}\lambda \\
\\
F_k^{'}\lambda & 0 & D_k^{'}\lambda & G_k^{'}(\lambda +\Lambda_k )
\end{array} \right),
\end{equation}\\
\\
and similarly $\;\;C_k^{'}=\frac{B_k^2}{\gamma k^2}(2k_x)\;,D_k^{'}=\frac{A_kB_k}{\gamma k^2}(2k_x)\;,F_k^{'}=\frac{A_kB_k}{\gamma k^2}(2ik_y)$,\\
\\
$G_k^{'}=\frac{A_k^2}{\gamma k^2}(2k_x)$.
\\
Using the following relation:
\begin{eqnarray}
\frac{1}{x -i\varepsilon}=Pr \frac{1}{x}+i\pi \delta (x)
\end{eqnarray}\\
Then the contribution of the K-point electrons on spin susceptibility, $\chi_{x x}$, is given by:
\begin{eqnarray}
\chi_{x x}(\omega)&=&\biggr(\sum_{\mu\nu}\int_{0}^{\infty}k dk\frac{2 \pi\lambda ^2\nu}{{e^{[\frac{1}{2}(-\nu\lambda +\mu\nu\Lambda_k)-\mu_{ch}]/K_B T}+1}}\Bigl [\frac{1}{\Lambda_k} Pr\frac{1}{\Lambda_k ^2-{(\hbar\omega)} ^2}\Bigl]\biggr)\nonumber \\
&&-2\pi\frac{\lambda}{\lambda ^2- {(\hbar\omega) }^2}\biggr\{\frac{1}{3}{(\frac{\pi K_BT}{\gamma})}^2+\frac{4\mu}{\gamma ^2}(\mu_{ch} +K_BT\ln (e^{-\frac{\mu_{ch}}{K_BT}}+1))\biggr\}\nonumber \\
&&+i \frac{\pi ^2}{4\gamma ^2 \hbar\omega} \biggr(\sum_{\mu\nu}\frac{\lambda ^2\nu}{{e^{[\frac{1}{2}(-\nu\lambda +\mu\nu\times\hbar\omega)-\mu_{ch}]/K_B T}+1}}\Bigl).
\end{eqnarray}
Where $\mu_{ch}$ is the chemical potential of the system.
\\
Finally,  both real ($\chi_{\mu\mu^{\prime}}$) and imaginary ($\chi^{''}_{\mu\mu^{\prime}}$) parts of spin susceptibility, can be obtained from the above expression.
\\
The calculations shows that circular symmetry around the Dirac points results in $\chi_{xx}=\chi_{yy}$ and $\chi_{xy}=\chi_{yx}=0$
\\
Similarly the $K^{'}$ point Hamiltonian reads:\\
\begin{equation}
\mathbf{H} = \left(
\begin{array}{cccc}
0 & 0 & -\gamma k_+ & - i\lambda\\
0 & 0 &0 & -\gamma k_+ \\
-\gamma k_-  & 0 & 0 & 0\\
i\lambda &-\gamma k_- & 0 & 0
\end{array} \right),
\end{equation}\\
The same eigenvalues are obtained for both of the Dirac points $K$ and $K^{'}$ while the eigenvectors are given by,
\begin{equation}
\mathbf{|\nu\mu >} =B_k\left(
\begin{array}{c}
\frac{i(\lambda - \Lambda_k)}{2\gamma k_+}\\
\\
1\\
\\
\nu \frac{ik_-}{k_+}\\
\\
\frac{\nu\lambda -\nu\mu \Lambda_k}{2\gamma k_+}
\end{array} \right);
\end{equation}
for $ \nu=\pm,\mu =+1$, and
\begin{equation}
\mathbf{|\nu\mu >} =A_k\left(
\begin{array}{c}
\frac{i(\lambda + \Lambda_k)}{2\gamma k_+}\\
\\
1\\
\\
\nu \frac{ik_-}{k_+}\\
\\
\frac{\nu\lambda -\nu\mu \Lambda_k}{2\gamma k_+}
\end{array} \right)
\end{equation}
for $ \nu=\pm,\mu =-1$.
\\
In this case, $s_x$ and $s_y$ on the basis of new $|\nu\mu>$ $K'$ states are obtained as follows:\\
\begin{equation}
\mathbf{s_x} = \left(
\begin{array}{cccc}
-C_k(\lambda - \Lambda_k ) & -D_k\lambda & 0 & F_k\Lambda_k\\
\\
-D_k\lambda & -G_k(\lambda +\Lambda_k ) & -F_k\Lambda_k & 0\\
\\
0 & F_k\Lambda_k & -C_k(\lambda - \Lambda_k ) & -D_k\lambda \\
\\
-F_k\Lambda_k & 0 & -D_k\lambda & -G_k(\lambda +\Lambda_k )
\end{array} \right),
\end{equation}\\
and
\begin{equation}
\mathbf{s_y} = \left(
\begin{array}{cccc}
-C_k^{'}(\lambda - \Lambda_k ) &-D_k^{'}\lambda & 0 &F_k^{'}\Lambda_k\\
\\
-D_k^{'}\lambda &-G_k^{'}(\lambda +\Lambda_k ) & -F_k^{'}\lambda & 0\\
\\
0 &F_k^{'}\Lambda_k &-C_k^{'}(\lambda - \Lambda_k ) & -D_k^{'}\lambda \\
\\
-F_k^{'}\lambda & 0 & -D_k^{'}\lambda &-G_k^{'}(\lambda +\Lambda_k )
\end{array} \right).
\end{equation}\\
We have performed the same calculations for spin susceptibility using the  $K^{'}$  representation of the spin operators where results show that the contribution of the both Dirac points in the spin susceptibility is the same at the level of the linear response regime.
\section{Results and discussion}
\begin{figure}[h]
\includegraphics{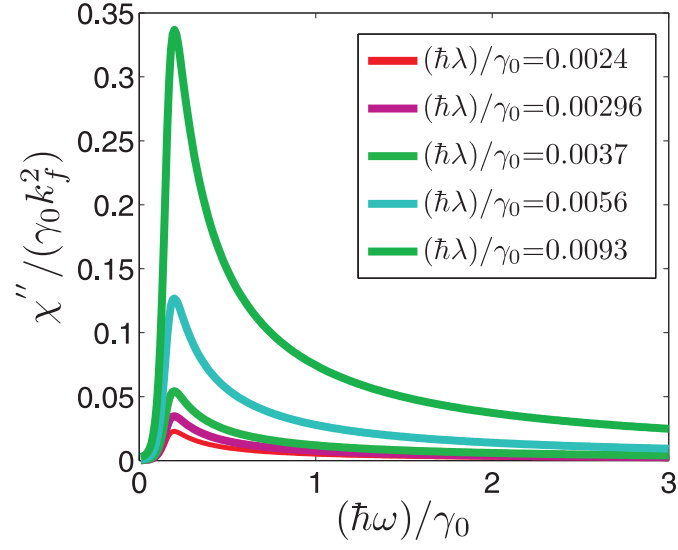}
 \caption{ Imaginary part of the spin susceptibility in terms of $\hbar\omega$.}
  \label{fig1}
\end{figure}

\begin{figure}[h]
\includegraphics{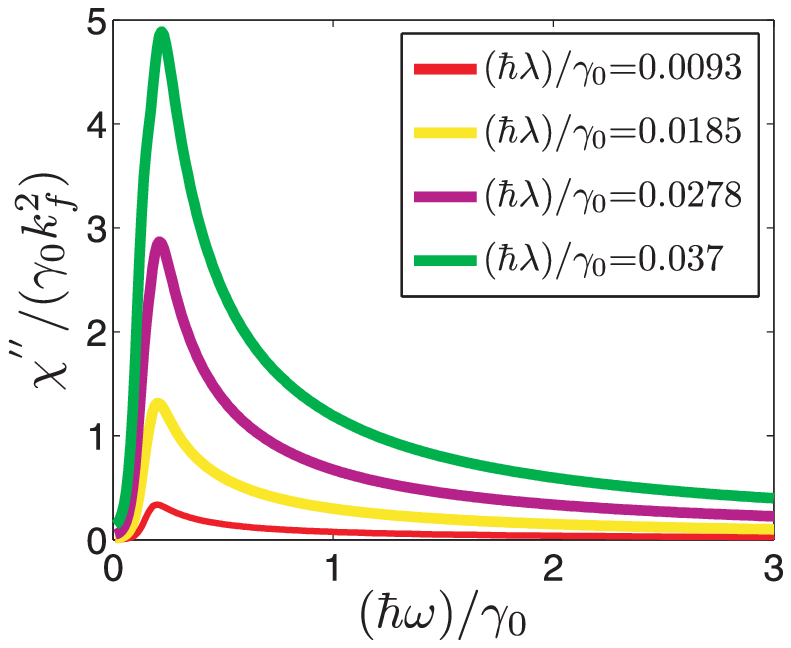}
 \caption{ Imaginary part of the spin susceptibility in terms of $\hbar\omega$.}
  \label{fig2}
\end{figure}

\begin{figure}\includegraphics{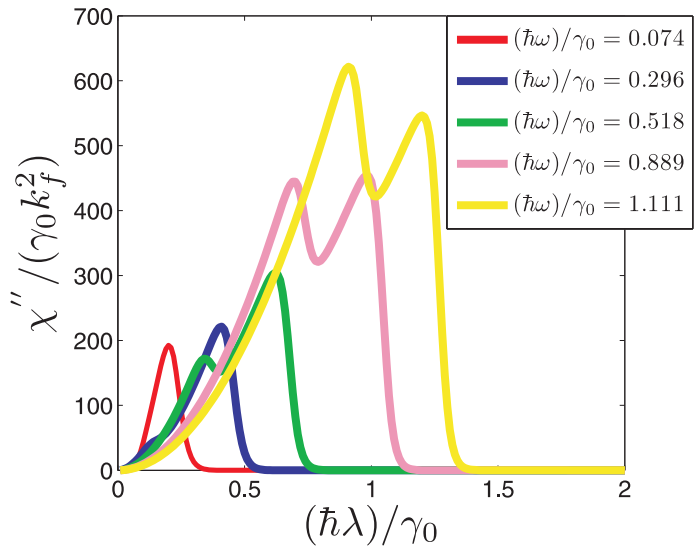}
 \caption{ Imaginary part of the spin susceptibility in terms of $\hbar\lambda$.}
  \label{fig3}
\end{figure}

\begin{figure}[h]
\includegraphics{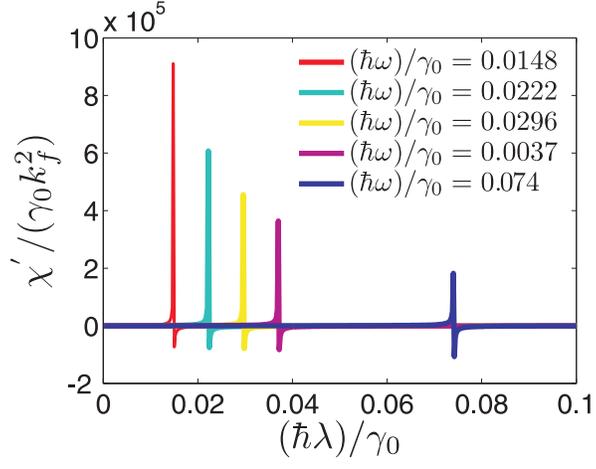}
\caption{Real part of the spin susceptibility in terms of $\hbar\lambda$.}
  \label{fig4}
\end{figure}

\begin{figure}[h]
\includegraphics{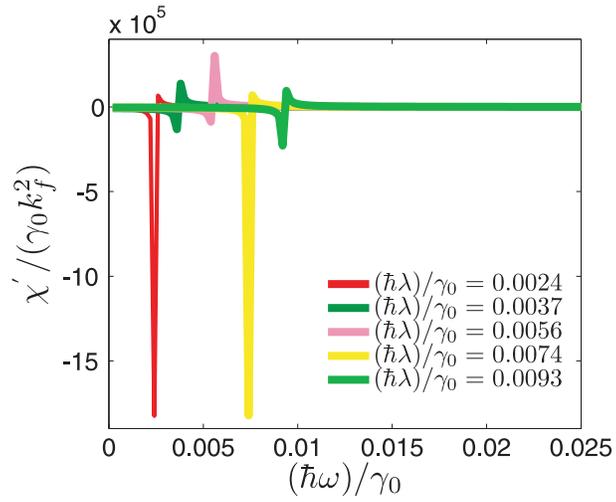}
 \caption{ Real part of the spin susceptibility in terms of $\hbar\omega$.}
  \label{fig5}
\end{figure}

For numerical calculations, the system parameters have been chosen as follows: $n=4\times 10^{16} ~m^{-2}$ is a typical density of electrons and holes in monolayer graphene \cite{Gong}, $k_f=3.54\times 10^8 m^{-1}$ is the wave vector at Fermi level,
and $K_BT=0.025$ eV describes the room temperature for the present calculations. Meanwhile the response function has been obtained for a experimentally available range of the Rashba coupling strength which has been reported up to $0.2$ eV.
\\
Results of the present work have been summarized in the figures Fig. \ref{fig1}-Fig. \ref{fig5}. Imaginary part of spin susceptibility has been depicted as a function of the frequency of the external field at different Rashba couplings as it was shown in figures Fig. \ref{fig1}-Fig. \ref{fig2}. Imaginary part of this response function measures the energy absorption of the system. This absorption arises as a result of the Zeeman coupling of the electron spins with the external magnetic field. As shown in these figures, increasing the Rashba interaction increases the optical absorption of the conduction electrons. The Rashba coupling increases the spin splitting of the band structure.
\\
It should be noted that unlike the two-dimensional electron gas systems, this type of spin splitting made by the Rashba interaction could be responsible for a small difference in the population of the spin bands of the carriers in mono-layer graphene. In the presence of the Rashba interaction the population of the spin bands is exactly the same in the two-dimensional electron gas system. In the other words, in this case, the different spin bands have the same form with a small displacement (determined by the Rashba coupling strength) in k-space, each state in one spin band has its counterpart with same energy in the opposite spin band. Therefore, a permanent magnetization of the moving spins cannot be generated by the Rashba coupling at equilibrium. In the mono-layer graphene the energy and therefore the population of the spin bands in the presence of the Rashba coupling are quite different. However, even in this case with different population of the spin bands the Rashba coupling cannot be responsible for magnetization of the carriers. This can be explained by the fact that the magnetization of the conduction electrons defined by $<S_i>=\Sigma_{\textbf{k},\mu\nu}s_{i\mu\nu}(\textbf{k})f(E_{k\mu\nu})$ exactly vanishes since the expectation value of spin operator $s_{i\mu\nu}(\textbf{k})=<k\mu\nu\mid s_i\mid k\mu\nu>$ is a odd function of \textbf{k} in either of the spin-bands while, the Fermi distribution function, $f(E_{k\mu\nu})$, is a even function in k-space.
Meanwhile this situation cannot be hold for out of equilibrium conditions where it was shown that non-equilibrium spin accumulation can be obtained in the two-dimensional electron gas \cite{Zhian}. This case can be realized, for example, in the presence of a current driving electric field. Accordingly, the magnetization and the filed absorption of the carriers could be effectively increased by increasing the Rashba coupling strength as shown in Fig. \ref{fig1}-Fig. \ref{fig2}.
\\
It is interesting to know that the amount of the frequency of maximum absorption is not significantly effected by the Rashba coupling strength. This could be expected regarding the fact that the magnitude of the Rashba coupling strength is significantly small in comparison with the hoping energy, therefore the corresponding change of the spin bands and spin splitting between the bands are really small.
These figures could be considered as the response of the system to the white field in which the absorption function has been represented for a wide range of the frequencies. Meanwhile the physical details of the response function can be inferred from the susceptibility of the system in the presence of monochromatic fields. Imaginary part of the spin susceptibility has been depicted as a function of the Rashba coupling in Fig. \ref{fig3}. As shown in this figure at low frequencies there is single extremum of the response function, however by increasing the frequency second maximum will be appeared (Fig. \ref{fig3}). Meanwhile the overall susceptibility increases by increasing the frequency. This could be understood if we consider that for a given frequency of the external field increasing the Rashba coupling results in displacement of spin bands and increasing the spin splitting. Therefore if the filed energy is high enough, two major and different types of absorption could be take place between the spin bands regarding the energy conservation issues in the absorption process.
\\
Singularities of the real response function corresponds to the Rashba coupling induced phase transition to a magnetically ordered carrier spins in which small magnetic field amplitude at  resonance frequency could result in significant magnetization of carriers. This phase characterizes by spontaneous magnetic ordering of conduction electrons (Fig. \ref{fig4}). Response function has a symmetric form for both $x$ and $y$ directions this can be understood as a result of the Dirac point approximation in which the anisotropic form of the energy dispersion relation of high energy states could be neglected at low energy Dirac points. As was also shown that, increasing the Rashba coupling strength changes the critical frequency of this magnetic phase transition (Fig. \ref{fig5}).
\section{Conclusion}
In this work spin-dependent features of the mono-layer graphene have been investigated in the presence of the substrate induced Rashba coupling. The numerical results have been obtained using the Kubo approach. Result of the current study demonstrates that, in the non-equilibrium regime the Rashba interaction has a central role in optical absorption and magnetic phase transition of the conduction spins.






\newpage
\newpage
Captions of the figures
\\
Fig 1:Imaginary part of the spin susceptibility in terms of $\hbar\omega$.
\\
Fig 2:Imaginary part of the spin susceptibility in terms of $\hbar\omega$.
\\
Fig 3:Imaginary part of the spin susceptibility in terms of $\hbar\lambda$.
\\
Fig 4:Real part of the spin susceptibility in terms of $\hbar\lambda$.
\\
Fig 5:Real part of the spin susceptibility in terms of $\hbar\omega$.
\newpage
\end{document}